\documentclass[12pt]{iopart}
\pdfoutput=1
\usepackage{xcolor}
\usepackage{graphicx}
\usepackage{subfigure}
\usepackage{lineno}

\usepackage{latexsym,amsmath,verbatim}
\usepackage{iopams}
\usepackage{hyperref}
\usepackage{siunitx}

\DeclareMathAlphabet{\Ibb}{U}{msb}{m}{n}

\newcommand{\diff}{\text{\rm d}}

\newcommand{\Bxi}    {\ensuremath{\boldsymbol\xi}}
\newcommand{\Beta}   {\ensuremath{\boldsymbol\eta}}

\newcommand{\Brho}{\ensuremath{\boldsymbol\rho}}

\newcommand{\Bb}{{\boldsymbol{\mathnormal b}}}

\newcommand{\Be}{{\boldsymbol{\mathnormal e}}}
\newcommand{\Bf}{{\boldsymbol{\mathnormal f}}}

\newcommand{\Bu}{{\boldsymbol{\mathnormal u}}}

\newcommand{\Bx}{{\boldsymbol{\mathnormal x}}}
\newcommand{\By}{{\boldsymbol{\mathnormal y}}}

\begin{document}

\title{Effects of disorder on deformation and failure of brittle porous materials}
\author{Jonas Ritter$^1$, Shucheta Shegufta$^{1,2}$ and Michael Zaiser$^1$}
\address{$^1$ Department of Materials Science, WW8-Materials Simulation, Friedrich-Alexander Universität Erlangen-Nürnberg (FAU), Dr.-Mack-str. 77, 90762 Fürth, Germany}
\address{$^2$ Central Institute for Scientific Computing (ZISC), Friedrich-Alexander Universität Erlangen-Nürnberg (FAU), Martensstrasse 5a, 91058 Erlangen, Germany}

\begin{abstract} 
The mechanical behavior of porous materials depends strongly on porosity and pore geometry, but also on morphological parameters characterizing the spatial arrangement of pores. Here we use bond-based peridynamics to study effects of disorder on the deformation and failure behavior of brittle porous solids both in the quasi-static limit and in case of dynamic loading scenarios. We show that structural disorder, which has a strong influence on stiffness, strength and toughness in the quasi-static limit, becomes less relevant under dynamic loading conditions. 
\end{abstract}

\noindent{\it Keywords\/}: Failure | Porous material | Disorder | Peridynamics

\section{Introduction}
\label{sec_Intro}

There exist an abundance of porous materials ranging from natural or engineered cellular materials \cite{gibson1989modelling} over geomaterials such as sandstone \cite{gal1998physical} or snow \cite{gerling2017measuring,sayers2021porosity} to building materials like concrete \cite{yaman2002active}. Inevitably the mechanical behavior and transport properties of such materials depend strongly on the morphology of their porous microstructure. Studies of linear  mechanical properties such as effective elastic moduli have traditionally focused on the role of porosity, drawing on results from the mechanics of composite materials (for an overview see e.g. \cite{yaman2002active}). At the same time it has been recognized that the arrangement of pores may have a significant impact even on linear mechanical properties such as the effective elastic moduli \cite{poutet1996effective}, and even more so on nonlinear properties characterizing the failure behavior. As demonstrated by Kun and co-workers using both discrete element simulations (DEM) and experimental data \cite{kun2013approach,pal2016record}, the approach to failure in disordered porous materials is characterized by precursor events in the form of avalanches with (truncated) power-law statistics, as typical for failure of disordered media \cite{alava2006statistical}. Laubie et. al. \cite{laubie2017stress} used lattice-element simulations to investigate stress transmission patterns in random porous media and their evolution in the run-up to failure. They showed that in such media the stress state in the approach to failure, characterized by strongly heterogeneous patterns resembling 'stress transmission chains', can be envisaged in analogy with the breakdown of force chains in the vicinity of the jamming transition of granular media. Huang et al. report similar observations in simulations using DEM and show increasing disorder in the arrangement of pores leads to a more rapid decrease of stiffness and failure strength with increasing porosity \cite{huang2021effect}.

Most simulation studies of the failure behavior of porous media envisage quasi-static loading scenarios where effects of inertia can be neglected. In the present investigation, we use peridynamic simulation to study the combined effects of porosity and disorder on the deformation and failure behavior of disordered porous materials, considering both quasi-static and dynamic loading 
scenarios. In \autoref{sec_Model_formulation}, we formulate the peridynamic simulation model and explain the method used for constructing porous microstructures with varying degrees of disorder. \autoref{sec_Sim_results} compiles simulation results for both quasi-static and dynamic loading, while a discussion of the results is given in \autoref{sec_Discussion}.

\section{Model formulation}
\label{sec_Model_formulation}

\subsection{Peridynamic model}
\label{subsec_Peri_model}

We use a bond-based peridynamic model as originally defined by Silling \cite{silling2010peridynamic}, considering two-dimensional systems which can be envisaged as the plane-stress deformation of porous thin sheets. Deformation is characterized by a displacement field $\Bu(\Bx) = \By(\Bx) - \Bx$ where $\By(\Bx)$ is the current location of the point with material coordinates $\Bx$.  The force balance equation for the point $\Bx$ is given by
\begin{equation}
	\rho(\Bx) \ddot{\Bu}(\Bx) = \int_{{\cal H}_{\Bx}} \Bf \left(\Bx,\Bx^{\prime} \right) \diff \Bx^{\prime} + \Bb(\Bx) ,
\end{equation}
where $\Bf \left( \Bx,\Bx^{\prime} \right)$ is the pair force between $\Bx^{\prime}$ and $\Bx$, $\Bb$ is a body force field, and interactions are restricted to a family ${\cal H}_{\Bx}$ which we take to be a circle of radius $\delta$, the so-called horizon, around $\Bx$, $\left\vert \Bx - \Bx^* \right\vert \le \delta \; \forall \; \Bx^* \in {\cal H}_{\Bx}$. 

The pair force is specified constitutively. We introduce the notations $\Bxi \left(\Bx,\Bx^{\prime} \right) = \Bx^{\prime} - \Bx$ and $\Beta \left(\Bx,\Bx^{\prime} \right) = \Bu \left(\Bx^{\prime} \right) - \Bu \left(\Bx \right)$. The pair force is then taken linearly proportional to the bond stretch $s$, and pointing in the direction of the vector $\Be_u$ connecting both points in the current configuration:
\begin{equation}
	\Bf \left(\Bx,\Bx^{\prime} \right)= c \left(\Bxi \right) s \Be_u \left(\Bxi,\Beta \right) \quad,\quad s = \frac{\left\vert \Beta + \Bxi \right\vert - \left\vert \Bxi \right\vert}{\Bxi}\quad,\quad \Be_u \left(\Bxi,\Beta \right) = \frac{\Beta + \Bxi}{ \left\vert \Beta+\Bxi \right\vert }
\end{equation}
Here $c(\Bxi)$ is the so-called bond micro-modulus which for an isotropic bulk material depends on the bond length $\xi = |\Bxi|$ only. In the following we use for simplicity a constant micro-modulus, $c(\xi)=c_0$. The bond energy can then be written in terms of the bond length $\xi$ and bond stretch $s$ as
\begin{equation}
	e \left(\Bx,\Bx^{\prime}\right) = 	e\left(\Bx^{\prime},\Bx\right) = \frac{c_0}{2} s^2 \xi.
\end{equation}

Elastic-brittle behavior is introduced by defining a critical bond stretch $s_{\rm c}$ in such a manner that, once $s > s_{\rm c}$, the bond fails: in this case, its micro-modulus is irreversibly set to zero. The elastic energy $\frac{c_0}{2} s_{\rm c}^2 \xi$ of the failed bond is then converted into defect energy $E_{\rm d}$; for our later considerations it is irrelevant whether the defect energy is considered as some kind of microscopically stored, non recoverable internal energy, or simply as heat. To avoid an artificial stiffening of the structures, a bond-filter prevents creation of bonds across narrow concave regions such as small pores. To this end, in the initial structure any bonds between material points that cross the pore space are deleted before the simulation begins.

We consider a 2D system with plane stress loading, which implies because of the central forces that Poisson's number $\nu = 1/3$. For discretization we use a square planar grid of collocation points with a spacing $d = \SI{0.25}{\mm}$, which also formally defines the 'thickness' of the system in the third dimension. The radius of the horizon is chosen as $\delta = \SI{0.75}{\mm}$. 

As to material parameters, the micro-modulus is chosen to  reproduce a Young's modulus $E = \SI{1000}{\giga \pascal}$, 
and the critical bond stretch is $s_{\rm c} = 0.001$, which sets the failure stress of a system under homogeneous uni-axial load to $\sigma_{\rm c} = \SI{1}{\mega \pascal}$. The density of the matrix material is taken as $\rho = \SI{1000}{\kilo \gram \per \cubic \metre}$, hence the longitudinal sound velocity in the matrix is $v_{\rm L} = \SI{1000}{\metre \per \second}$. 

\subsection{Geometry generation}
\label{subsec_Geo_generation}

\begin{figure}[tbh]
	\begin{center}
		\vspace{-0cm}
		\includegraphics[width=\textwidth]{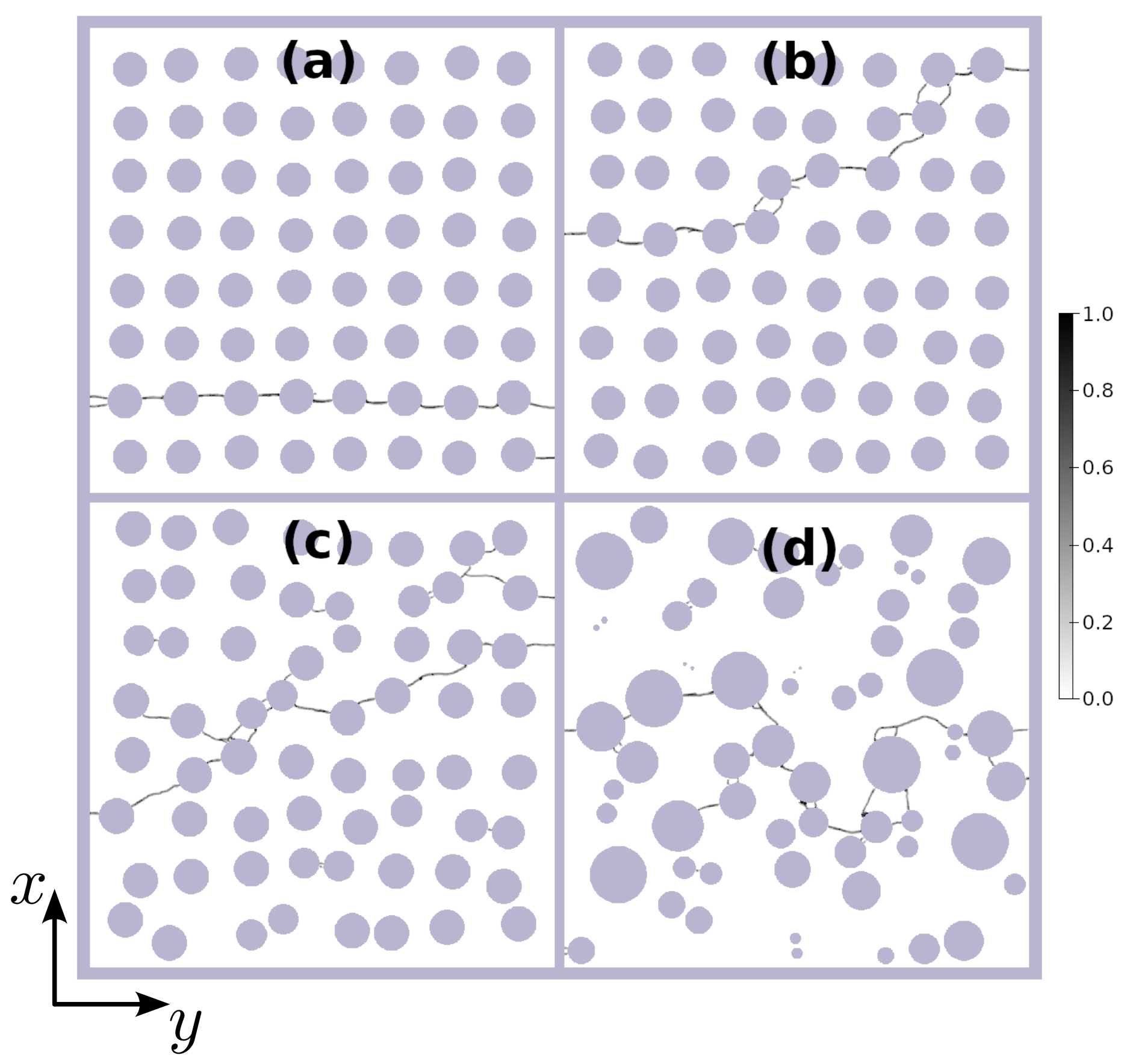}
		\caption{Structures of different disorder with corresponding failure patterns, porosity $\phi = 0.3$ for all patterns, for all simulations $L = \SI{200}{\mm}$, $n=8$, $\delta = \SI{0.75}{\mm}$, material parameters $E = \SI{1000}{\mega \pascal}$, $\nu = 1/3$ (plane stress); (a) $\eta = 0.06$, (b) $\eta = 0.12$, (c) $\eta = 0.24$, (d) $\eta = 0.48$; the colorscale indicates the damage parameter $\varphi$ defined as fraction of broken bonds in the horizon of a collocation point.}
		\label{fig:poresamples}
	\end{center}
\end{figure}

\begin{figure}[tbh]
	\begin{center}
		\vspace{-0cm}
		\includegraphics[width=\textwidth]{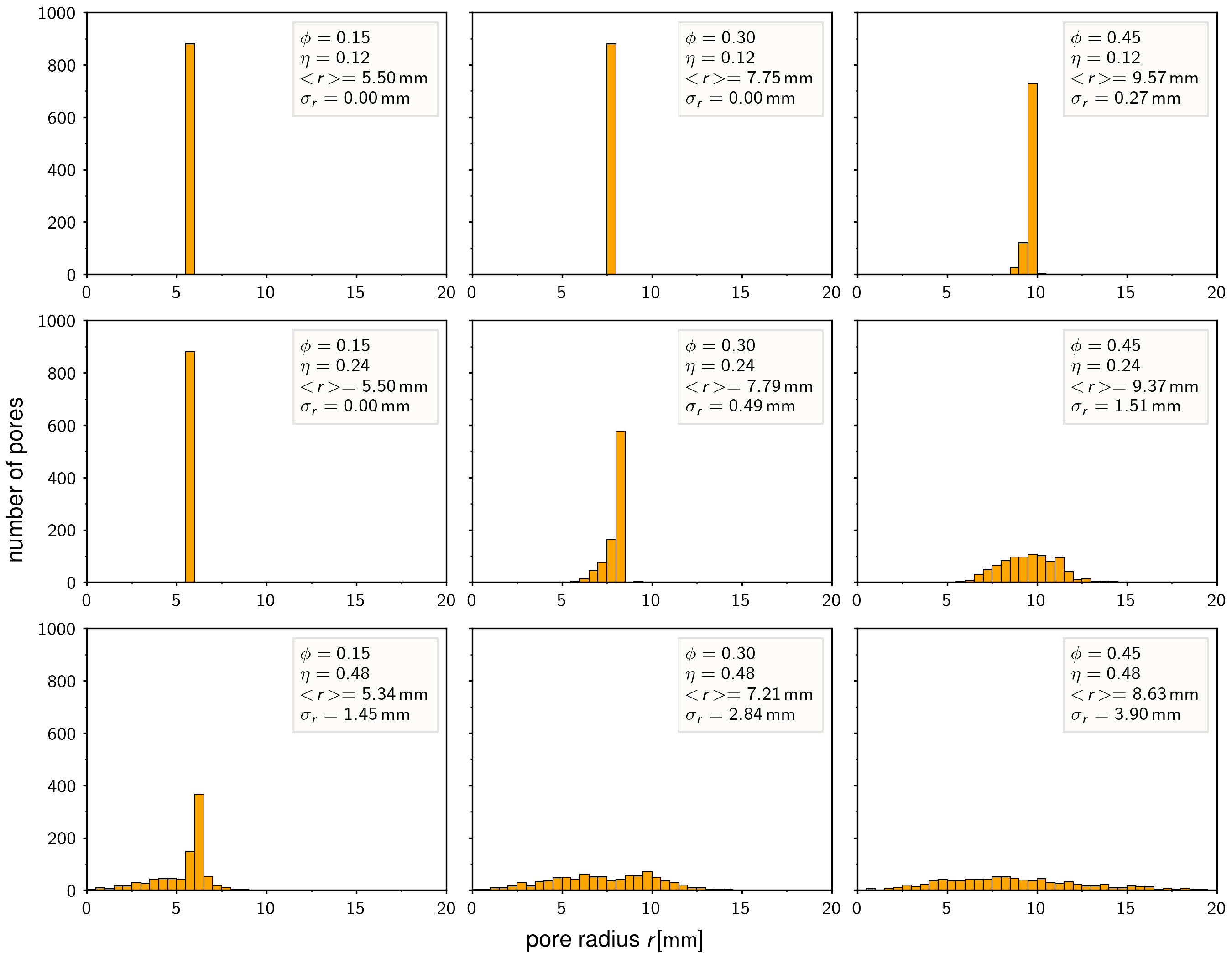}
		\caption{Histograms of the pore size distriution for different values of porosity $\phi$ and disorder $\eta$, all histograms correspond to $N=880$ simulated pores, porosity: $\phi = 0.15$ (left column), $\phi = 0.3$ (middle column), $\phi = 0.45$ (right column); disorder parameter $\eta = 0.12$ (top row), $\eta = 0.24$ (middle row), (d) $\eta = 0.48$ (bottom row).}
		\label{fig:poresize}
	\end{center}
\end{figure}

We consider two-dimensional square samples of size $L\times L$ containing spherical pores whose arrangement and size exhibit random variation while the overall porosity $\phi$ adheres to a prescribed value. In constructing our model structures, we use a variation of the method proposed by \cite{liebenstein2018size} for generating random cellular structures. We start with a regular square grid of $n\times n$ 'seeds'; the lattice constant $l = L/n$ of this grid defines our mean pore spacing, which we keep fixed at $l = \SI{25}{\mm}$. Next, each seed $i$ is displaced by a random vector $\Brho_i = (\rho_i,\theta_i)$ where the displacement directions $\theta$ are independent random variables that are equi-distributed over the interval $0 \le \theta < 2\pi$, and the displacement distances $\rho_i$ are independent random variables that are equi-distributed over the interval $0 \le \rho  \le \eta l$. For construction of the seed pattern the system is continued periodically in all spatial directions, and seeds that are displaced out of the original system are projected back from the periodic image to which they are displaced. The parameter $\eta$ characterizes the disorder of the resulting seed arrangement; for $\eta = 0$ we retain the initial periodic grid, for $\eta \rightarrow \infty$ seeds are placed in a completely random manner following a 2D Poisson point process. Finally, from each seed we 'grow' a spherical pore according to the following rules:

\begin{itemize}    
	\item All pores grow initially at equal rates.
	\item If the distance between two pores is less than $2 \delta$, both pores cease to grow.
	\item If the distance between a pore and the surface is less than $\delta$, it ceases to grow.
	\item The growth of all other pores stops once porosity reaches the pre-defined value $\phi$.
\end{itemize} 

This method imposes a fixed (spherical) pore shape, while the pore size distribution and pore arrangement depend on the disorder parameter $\eta$. Examples are shown in \autoref{fig:poresamples}. The width of the pore size distribution increases with increasing disorder parameter $\eta$ and also with increasing porosity $\phi$, as shown by the pore size histograms in \autoref{fig:poresize}. Alternative methods of generating random porous structures may be based on random fractals such as Appollonian packings \cite{anishchik1995three} or on level-cutting random fields, see e.g. the use of thresholded Gaussian random fields for modelling snow microstructures \cite{blatny2021computational}. For this study we considered structures of three different sizes ranging from $L = \SI{50}{\mm}$ over \SI{100}{\mm} up to \SI{200}{\mm}. For the smallest size 20 samples and for the two others sizes each 10 samples were generated.

\subsection{Boundary conditions}
\label{subsec_Boundary_conditions}

We introduce a Cartesian coordinate system aligned with the edges of the sample, with the origin in one corner, such that the sample occupies the domain $0 \le x \le L, 0 \le y \le L$. Macroscopically uni-axial loading is imposed in $x$ direction by fixing a layer of width $-\delta \le x \le 0$ at one side of the sample ('support layer') while the opposite layer $L \le x \le L+\delta$ is displaced in $x$ direction by $\Delta L = \varepsilon_{xx} L$ ('displaced layer'). Both layers are free to cross-contract in $y$ direction, and no boundary conditions are imposed on the other surfaces. We note that, in order to avoid surface artefacts related to the so-called peridynamic surface effect (for discussion, see \cite{ritter2022energetically}), the size of the support and displaced layer are of the horizon size.

Two loading protocols are considered: 

(i) In {\em quasi-static loading} we neglect inertia and carry out sequential energy minimization steps. Thus, the applied global strain is increased in small steps $\Delta \varepsilon_{xx} = \gamma s_{\rm c}$ where we take $\gamma = 0.01$. After each strain increment, force equilibrium is sought by minimizing the energy of the system while removing any bonds that become stretched beyond the critical level $s_{\rm c}$. The global load is then computed in terms of the total reaction force on the support layer (or, equivalently, on the displaced layer, since force equilibrium requires both forces to be equal). The simulation is terminated when no equilibrium configuration can be found, indicating failure. Quasi-static loading has been studied both in tension and in compression.

(ii) In {\em dynamic loading} we impose a time-dependent displacement on a surface layer at $L \le x \le L+\delta$ and solve Eq. (1) at the collocation points while the opposing surface is constrained to remain at $x = 0$. The displacement follows the protocol
\begin{equation}
	\partial_t \Delta L =
		\begin{cases}
			v_0 \frac{t}{t_{\rm a}} & \text{if } t\le t_a,\\
			v_0 & \text{if } t> t_a.
		\end{cases}
\end{equation}
The initial acceleration time $t_{\rm a}$ is introduced to avoid artefacts arising due to infinite acceleration if $t_{\rm a} \to 0$, it is chosen to be a fraction of the time needed for a compression wave to travel through the sample: $t_{\rm a} = \zeta (L/v_{\rm L})$ where $\zeta < 1$ is chosen such that the results are insensitive to the choice of acceleration time. After acceleration, the imposed global strain rate takes the constant value $\dot{\varepsilon}_{xx} = v_0/L$. During loading we separately record the stresses on the displaced layer $L \le x \le L+\delta$ and the support layer $-\delta \le x \le 0$. We also record the total accumulated damage and kinetic energy of the system. 
In dynamic loading simulations, we focus on compression where the differences between static and dynamic loading are most pronounced. To account for interaction between fragments during compression we utilize a global penalty type contact formulation.

\section{Simulation results}
\label{sec_Sim_results} 

\subsection{Quasi-static loading in tension}
\label{subsec_Quasistatic_loading}

\subsubsection{Elastic modulus}
\label{subsubsec_Elasitc_modulus}

\begin{figure}[tbh]
	\begin{center}
		\vspace{-0cm}
		\includegraphics[width=16cm]{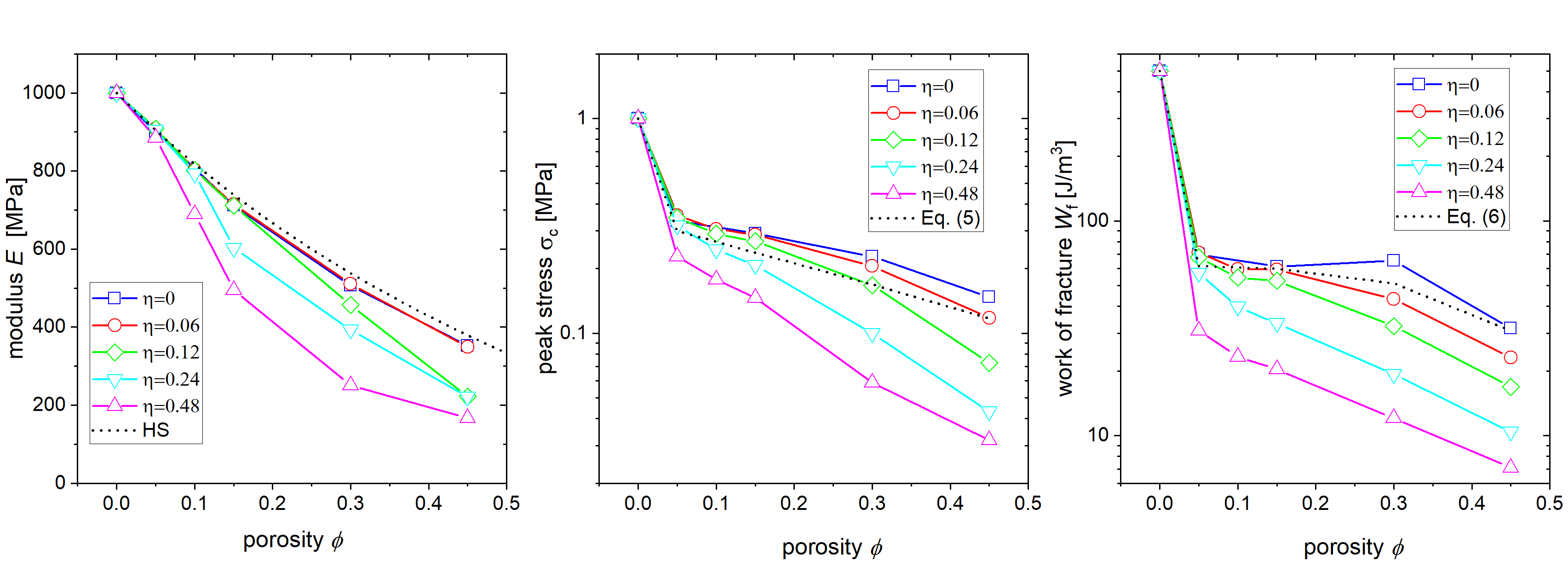}
		\caption{Dependency of mechanical properties on porosity, for different values of the disorder parameter $\eta$ as given in the legend; left: elastic modulus $E$, the blue curve indicates the upper Hashin-Shtrikman bound; center: peak stress $\sigma_{\rm c}$, the blue curve indicates the expected behavior for a single void in an effective medium constituted by the porous material, Eq. (5); right: work of failure $W_{\rm f}$, the blue line indicates the prediction for perfect elastic-brittle behavior, Eq. (6).}
		\label{fig:quasistatic}
	\end{center}
\end{figure}

We first look at the dependency of elastic modulus (Young's modulus $E$) on porosity and disorder. Since the system is allowed to freely cross-contract, in case of a homogeneous system the stress state is purely uni-axial and $E = \sigma_{xx}/\epsilon_{xx}$. We adopt the same definition for the effective modulus of a porous system. \autoref{fig:quasistatic}, left, shows thus defined elastic moduli for systems of different porosity and disorder. In order to assess geometry specific effects, it is instructive to compare these moduli with results of Hashin and Shtrikman \cite{hashin1962variational}, who provided rigorous upper and lower bounds on the effective properties of composites based on the volume fractions and properties of the constituent materials (here: matrix material and void space, with porosity being identical with the volume fraction of void space) alone, without any consideration of geometry. Since for a porous material the lower Hashin-Shtrikman bound of the elastic modulus is exactly zero, we consider as reference the porosity-dependent upper Hashin-Shtrikman bound. At low disorder ($\eta = 0$ and $\eta = 0.06$), the decrease of modulus with increasing porosity follows closely the theoretical limit curve. In the limit $\phi \to 0$ the effective elastic moduli of the porous samples smoothly approach the modulus of the bulk material. 
With increasing disorder, the knock-down effect of porosity on elastic stiffness becomes more pronounced, in particular in the regime of high porosities where elastic moduli of disordered systems fall significantly below the Hashin-Shtrikman bound. 

\subsubsection{Peak strength and work of failure}
\label{subsubsec_Peak_strength}

Looking at peak strength, the impact of porosity is much more dramatic (\autoref{fig:quasistatic}, center). Even for the smallest porosities, the peak strength is only about one third of the strength of the pore-free bulk material and with increasing porosity the values decrease further. This decrease again becomes more pronounced with increasing disorder. 

This behavior is readily understood if we recall a result by Kirsch \cite{kirsch1898theorie} which states that, around a circular hole in an infinite plate subject to uni-axial stress $\sigma_{xx}$, the peak stress $\sigma_{\rm p}$ = 3$\sigma_{xx}$ reached at the surface of the hole is three times the axial stress irrespective of the size of the hole. The locations of this stress concentration are on the maximum cross section perpendicular to the loading direction. We may thus interpret the observed behavior by assuming that failure occurs by nucleation of damage at these locations, while the influence of the other voids in the system can be accounted for by a reduced effective elastic modulus $E(\phi)$. We thus treat the medium with multiple voids as equivalent to a single void in an effective medium of reduced modulus. In an isostrain approximation, the failure strain of the effective medium is the same as that of the matrix material, and local failure then leads to crack nucleation at the locus of maximum stress concentration once $\sigma_{\rm p} = E(\phi) \epsilon_{\rm c}$. With $\sigma_{\rm p}$ = 3$\sigma_{xx}$, this leads to the simple relation
\begin{equation}
	\sigma_{xx,\rm c} = \frac{1}{3}E(\phi) \epsilon_{\rm c}. 
\end{equation}
As can be seen from \autoref{fig:quasistatic}, center, this equation -- which implies a mean-field treatment of the effect of other voids, which enter the failure criterion only through the reduction of elastic modulus with increasing porosity $\phi$ -- predicts the behavior at low disorder reasonably well. At higher disorder, the stress concentrations created by nearby voids may act in a synergetic manner, leading to 'weak spots' where the ligaments between voids have below-average thickness, and hence to a stronger strength reduction than predicted in the single-void mean-field approximation. 

A similar consideration can also be made for the work of failure. In an effective medium approximation and assuming that failure occurs immediately after the first crack nucleates at a void, the work of failure is simply the elastic energy stored at the critical stress, hence
\begin{equation}
	E_{rm f} = \frac{1}{2E} \sigma_{xx,\rm c}^2. 
\end{equation}
Again, this describes the work of failure at low disorder reasonably well but over-estimates the knockdown effect of increased disorder (\autoref{fig:quasistatic}, right.)

\subsubsection{Size dependence of mechanical properties}
\label{subsubsec_Size_dep}

\begin{figure}[tbh]
	\begin{center}
		\vspace{-0cm}
		\includegraphics[width=16cm]{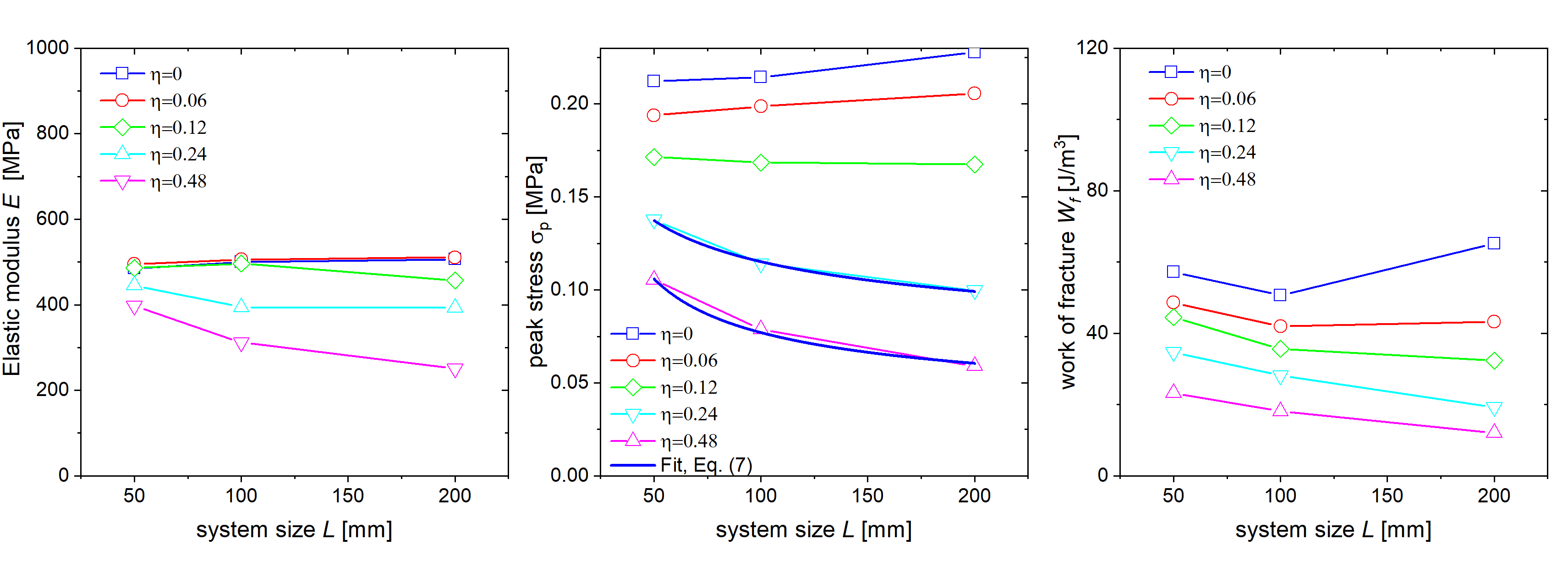}
		\caption{Dependency of mechanical properties on size for samples of \SI{30}{\%} porosity and different values of the disorder parameter $\eta$ as given in the legend; left: elastic modulus $E$; center: peak stress $\sigma_{\rm p}$, blue line: fit of high-disorder data by Eq.(7); right: work of failure $W_{\rm f}$.}
		\label{fig:sizeeffect}
	\end{center}
\end{figure}

In peridynamic simulations of disordered porous materials, size dependent mechanical behavior can arise from two distinct sources: First, larger samples have less outer surface, which means that the influence of the peridynamic surface effects \cite{ritter2022energetically} decreases, leading to  stiffening and strengthening of larger samples. Second, in disordered samples both elasticity and failure may be controlled by 'weakest links' in terms of cross-sections of locally enhanced porosity, giving rise to statistical size effects, as weakest links are, on statistical average, expected to be weaker in larger samples. To investigate these counter vailing tendencies, we perform simulations of samples of equal pore density but different geometrical size, ranging from $L = \SI{50}{\mm}$ to $L = \SI{200}{\mm}$ with a fixed average pore spacing of $\SI{25}{\mm}$. Results for \SI{30}{\%} porosity are compiled in \autoref{fig:sizeeffect}. 

It can be seen that indeed the size dependency of mechanical properties changes with increasing disorder. For low disorder, we see almost no size dependence of the elastic modulus, and a weak increase of strength and work-of-failure with increasing system size that can be attributed to the decreasing importance of boundary effects. Conversely, at high disorder, elastic modulus, peak stress and work of failure all decrease with increasing system size, indicating the relevance of statistical size effects. The decrease of peak stress with system size can, in line with findings for other disordered systems \cite{alava2009size,taloni2018size}, be described by a proportionality to the inverse logarithm of system size, 
\begin{equation}
	\sigma_{\rm p}=\frac{A}{B+\ln(L)}
\end{equation}
where the parameters $A$ and $B$ need to be fitted to the data.

\subsection{Dynamic loading in compression}
\label{subsec_Dyn_loading}

\begin{figure}[tbh]
	\begin{center}
		\vspace{-0cm}
		\includegraphics[width=6.8cm]{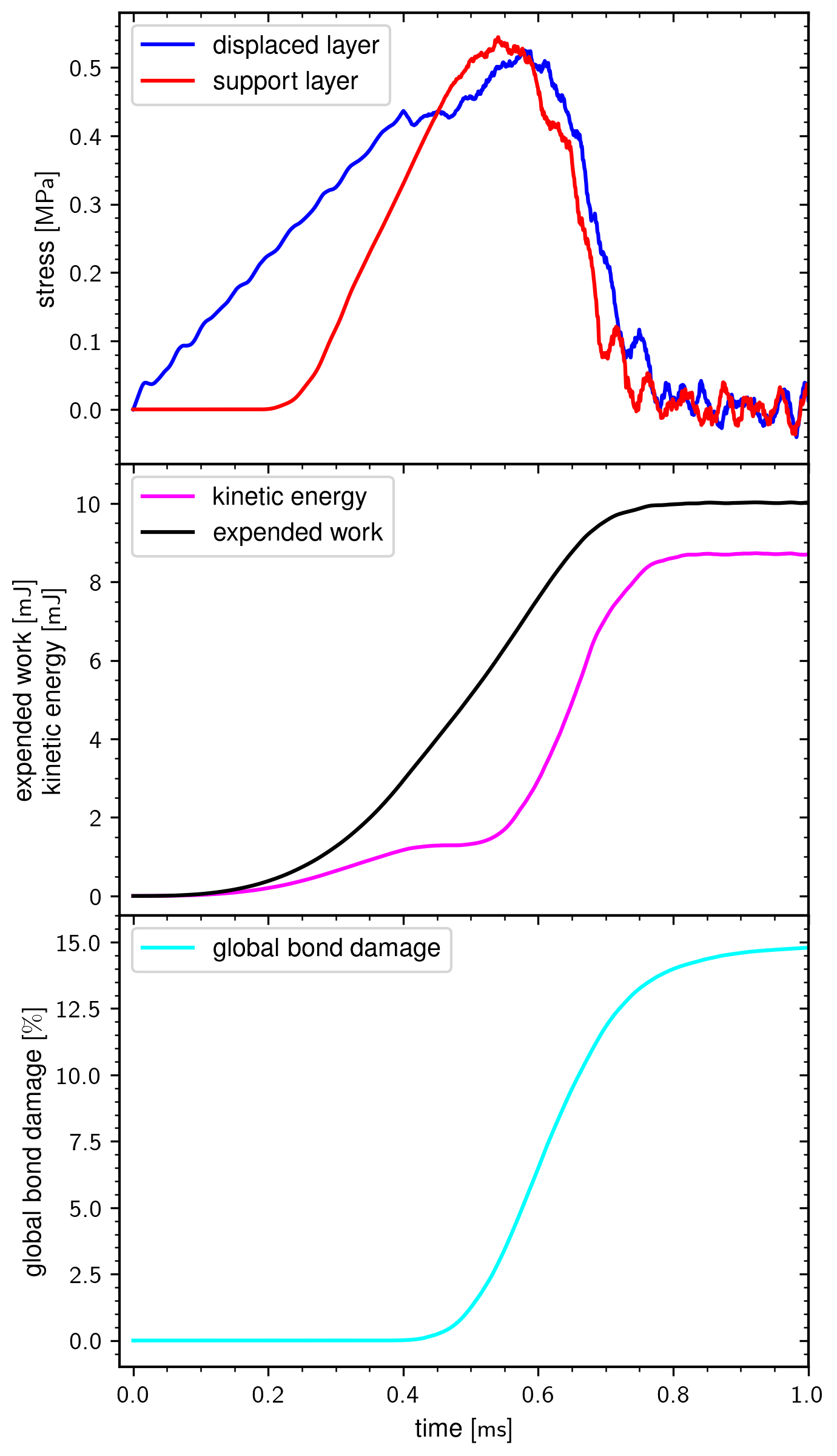}
		\caption{Dynamic loading of a sample with $L = \SI{200}{\mm}$ (porosity \SI{45}{\%}, $\eta = 0.06$) at a velocity of $v_0 = \SI{1}{\metre \per \second}$ with acceleration time $t_{\rm a} = 0.4$ ms; top: stress vs time at the opposite sides of the sample (displaced layer and support layer); center: time evolution of expended work and kinetic energy; bottom: time evolution of the fraction of broken bonds.}
		\label{fig:dynamicsample}
	\end{center}
\end{figure}

In dynamic loading situations, kinetic energy contributions and acceleration forces are non negligible. The stress state is inherently non-uniform even if the samples are homogeneous. A typical dynamic loading scenario is depicted in \autoref{fig:dynamicsample}, showing the loading of a sample of size $L = \SI{200}{\mm}$ at a velocity of $v_0 = \SI{1}{\metre \per \second}$ with an initial acceleration period $t_{\rm a} = 0.4$ ms. 
The load signals at the displaced layer and support layer are initially different: Until about \SI{0.2}{\milli \second} (run-time of the compressive wave through the sample), the load at the support is zero but then rapidly rises. The load on the displaced layer, on the other hand, increases continually until, from about $t_{\rm a} = \SI{0.4}{\milli \second}$ onwards, both signals are almost identical, indicating that stress equilibrium is reached. In particular, the peak amplitude of both stress pulses is almost identical and can be used as a signature for the stress required for inducing dynamic fragmentation. Damage as quantified by the fraction of broken bonds (i.e., number of broken bonds divided by the initial number of all bonds connecting collocation points in the discretized version of the peridynamic system) then accumulates for about \SI{0.4}{\milli \second} while the load drops. From about \SI{0.8}{\milli \second} onwards, the load oscillates around zero and the expended work, kinetic energy and accumulated damage almost saturate. At this stage, the fracture and fragmentation process is essentially over as the damage level reached corresponds to over \SI{80}{\%} of the damage accrued over a total simulation time of \SI{30}{\milli \second} and the kinetic energy level to almost \SI{70}{\%} of the total kinetic energy imparted on the sample. Generally speaking, stable values of the total damage, kinetic energy and work of failure are obtained by considering the situation at the point in time when the stress signal on the support (which for obvious reasons tends to lag behind the stress on the displaced layer) passes for the first time through zero.

We note that the integral signatures of the process (expended work, kinetic energy, accumulated damage and peak stress) are insensitive to the choice of acceleration time, which we have varied by a factor of $4$. Leaving out the acceleration phase (in fact, imparting infinite acceleration on the displaced layer) leads however to artefacts in form of a extremely high stress peak on the displaced layer shortly after start of the simulation, and makes it thus impossible to meaningfully determine a fracture or fragmentation stress level. 

In the following we always use an acceleration time $t_{\rm a} = \SI{0.4}{\milli \second}$. We record the following integral signatures of the failure process : Peak stress on the displaced layer, peak stress on the support, accrued damage, total expended work, and total kinetic energy, all of which are determined at the moment when the support stress for the first time passes through zero. We consider systems of size $L = \SI{200}{\mm}$, with displacement velocities in the range \SI{0.02}{\metre \per \second} $\le v_0 \le $ \SI{5}{\metre \per \second}, corresponding to nominal strain rates $\SI{0.1}{\per \second} \le \dot{\epsilon} \le \SI{25}{\per \second}$. In addition, for each set of parameters (porosity, disorder) quasi-static compression tests were performed as reference for comparison.

\begin{figure}[tbh]
	\begin{center}
		\vspace{-0cm}
		\includegraphics[width=16cm]{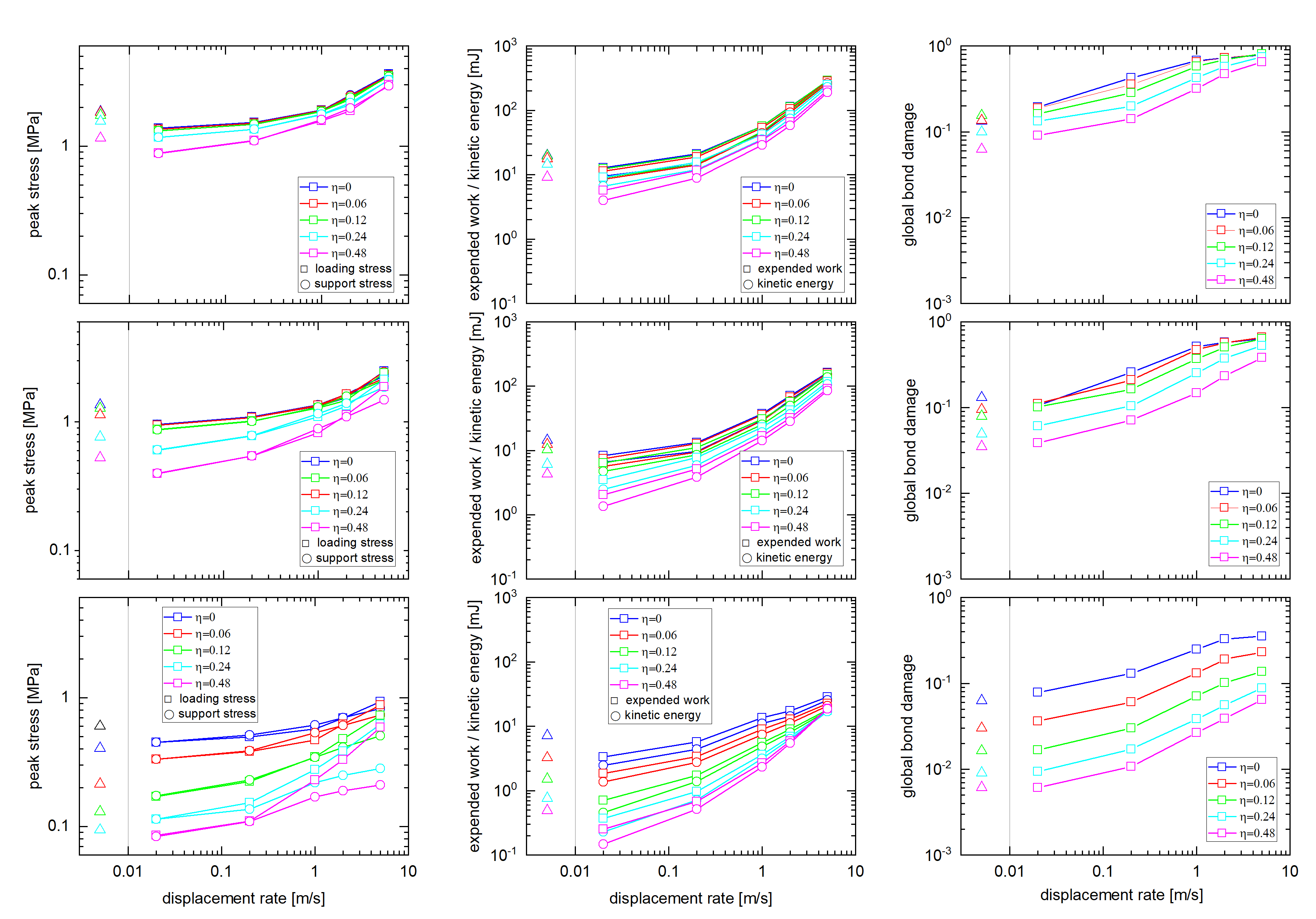}
		\caption{Dependency of mechanical properties on displacement velocity in dynamic testing; the graphs in the top row show the situation at \SI{5}{\%} porosity and different values of the disorder parameter $\eta$ as given in the legend; the middle row corresponds to \SI{15}{\%} porosity, and the bottom row to \SI{45}{\%} porosity; left: peak stress (square symbols: at displaced layer, circular symbols: at support), center: expended work (square symbols) and total kinetic energy (circular symbols) at end of fragmentation stage, right: total damage (fraction of broken bonds) at the end of the fragmentation stage; in all figures the triangular symbols located at the left outside the logarithmic $x$-axis scale represent corresponding data from quasi-static compression simulations.}
		\label{fig:dynamicfail}
	\end{center}
\end{figure}

Results are compiled in \autoref{fig:dynamicfail}, which shows the integral signatures of the failure process for different porosities (porosity increasing from top to bottom). The general trend follows the observations from the quasi-static simulations. i.e., disorder has a negative impact on peak stress and work of failure, which becomes more pronounced with increasing porosity. The same 
is true for the damage that accumulates during the fragmentation process. It is evident that increasing porosity implies that less bonds need to be broken in order to disconnect the structure, and it is interesting to note that disorder has a comparable effect as fragmentation proceeds in a less homogeneous manner in disordered structures (see animations in supplementary material). 

\begin{figure}[tbh]
	\begin{center}
		\vspace{-0cm}
		\includegraphics[width=16cm]{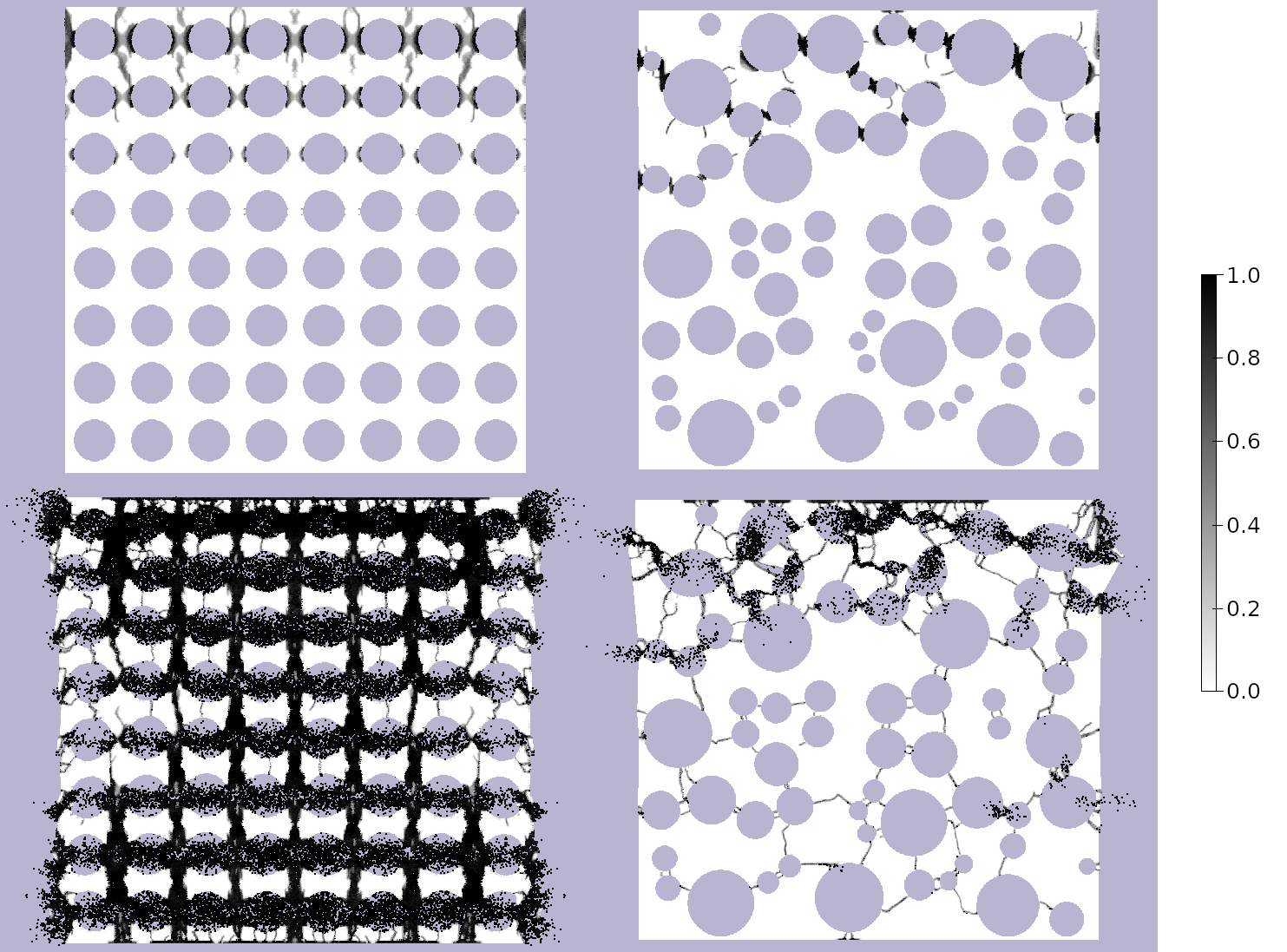}
		\caption{Fracture and fragmentation pattern for a regular ($\eta = 0$, left column) and disordered ($\eta = 0.48$, right column) structure with a porosity of \SI{45}{\%} compressed by a rate of $\dot{\epsilon} =  \SI{25}{\per \second}$. The top row shows the fracture pattern at the time $t = \SI{0.2}{\milli \second}$ and the bottom row the fragmented structure at $t = \SI{2}{\milli \second}$; the colorscale indicates the damage parameter $\varphi$ defined as fraction of broken bonds in the horizon of a collocation point.}
		\label{fig:dynamicfailFracture}
	\end{center}
\end{figure}

In most simulations, the stress peaks at the displaced layer and at the support are of comparable magnitude, and indeed the simulations proceed in approximate stress equilibrium once the initial stage of wave propagation is over. However, a notable exception is provided by structures that are both highly disordered and highly porous (see the curves pertaining to high disorder parameters 
$\eta$ in the bottom left graph of \autoref{fig:dynamicfail}). In these structures, deformation at the highest strain rates proceeds in an intrinsically heterogeneous manner where weak fracture pathways near the displaced layer are activated and lead to early detachment and rapid acceleration of large chunks of material, see \autoref{fig:dynamicfailFracture}. This leads to a significant reduction of the stress wave magnitude transmitted to lower layers which, at the highest degrees of disorder, may be up to a factor $3$ less in peak amplitude than the incident stress wave. Thus, the idea that 'disorder is good for you' \cite{tuzes2017disorder} holds also for the effects of disorder on shock absorption. 

At the same time it may be observed that the effects of disorder tend to decrease with increasing displacement rate, as manifest by the general tendency in \autoref{fig:dynamicfail} for curves pertaining to different disorder parameters to mutually converge towards high displacement rates (the mentioned stress anomaly being a notable exception). Indeed, at the highest displacement rates the major factor controlling both the peak stress and the work of failure is provided by the force and work need to rapidly accelerate the impacted material. At the same time, geometrical factors (disorder) of the structural arrangement become of relatively less importance. 

We finally note that quasi-static simulations consistently result in values of the peak stress and expended energy that are slightly higher than those in the low-rate limit of dynamic simulations. This at first glance surprising behavior is readily explained if we note that, in a quasi-static simulation, any energy that is released subsequent to local fracture is instantaneously dissipated as inertial effects are completely absent. In slow dynamic simulations, on the other hand, the released energy is converted into kinetic energy and only gradually transferred away from the locus of the initial fracture. This localized kinetic energy can, via the associated inertial forces, help to cause secondary fractures and thus facilitate dynamic fracture propagation, which as a consequence becomes possible at a reduced stress level. Similar observations have been reported in molecular dynamics simulations of disordered graphene sheets, where quasi-static ‘molecular mechanics’ predicts higher fracture stresses than zero-temperature molecular dynamics with a finite response time of the thermostat: in the second case the kinetic energy release from previous bond fractures facilitates bond fracturing at the crack tip and enables crack propagation at a reduced stress level \cite{nasiri2016rupture}.

\section{Discussion and conclusions}
\label{sec_Discussion}

We have studied the effects of disorder in quasi-static and dynamic failure of disordered porous materials. Our findings regarding quasi-static tensile failure fit into the general paradigm of load-driven failure of disordered materials. Structural disorder leads to 'weak spots' which have a strong impact on the elastic modulus, as manifested by moduli that fall significantly below the porosity-dependent Hashin-Shtrikman bounds. An even stronger effect is found for the disorder dependence of peak stress and work-of-failure, which cannot be explained in terms of the decreasing modulus alone and indicates statistical effects in the sense of weakest-link behavior. This viewpoint is corroborated by the size dependence of fracture strength, which shows a logarithmic size effect as often observed in studies of failure of disordered materials. 

Compared to quasi-static fracture, dynamic fracture of disordered materials has received only limited attention in the literature. Our findings indicate that the impact of disorder tends to decrease with
increasing deformation velocity, as acceleration and inertial effects become more and more pronounced while structural aspects become, relatively speaking, less important. While this tendency is expected, we however note some remarkable, and interesting, deviations from the general trend when it comes to stress transmission between the impacted layer and the support of the samples. In this case, disorder seems to fulfill an important role in deflecting the transmission of linear momentum to the support of the sample, such that the stress wave transmitted to the support is much attenuated, with obvious beneficial implications in view of impact protection. More studies may be needed to explore this feature, which is most pronounced in the regime of high porosities and high disorder. Such studies will need to expand the investigation from 2D to realistic 3D materials, which implies a very considerable additional computational cost -- but we are optimistic that the qualitative results of the present investigation will carry over to the 3D case.


\section*{Declarations}

\subsection{Data availability}

Datasets generated and analyzed during this study are published and available on Zenodo, \url{https://doi.org/10.5281/zenodo.7705091}.

\subsection{Competing interests}

The authors declare that they have no competing interests.

\subsection{Author's contributions}

J.R. and S.S. performed peridynamic simulations and data analysis, M.Z. designed the study and drafted the manuscript. The manuscript was edited and approved jointly by all authors.

\subsection{Funding}

This work was funded by the Deutsche Forschungsgemeinschaft (DFG, German Research Foundation) - 377472739/GRK 2423/1-2019. The authors gratefully acknowledge this support.

\section*{Acknowledgements}  
  
\section*{References}
\bibliographystyle{iopart-num}
\bibliography{periref}

\end{document}